\documentclass[12pt]{article}
\usepackage{a4}
\usepackage{epsfig}

\newlength{\abstwidth}
\def\be{\begin{equation}} 
\def\ee{\end{equation}} 
\def\bq{\begin{equation}} 
\def\eq{\end{equation}} 
\def\bqa{\begin{eqnarray}} 
\def\eqa{\end{eqnarray}} 
\def\lsim{\roughly<}
\def\gsim{\roughly>}

\begin{document}

\def\lsim{\mathrel{\rlap{\lower4pt\hbox{\hskip1pt$\sim$}}
    \raise1pt\hbox{$<$}}}         %less than or approx. symbol
\def\gsim{\mathrel{\rlap{\lower4pt\hbox{\hskip1pt$\sim$}}
    \raise1pt\hbox{$>$}}}         %greater than or approx. symbol

\pagestyle{empty}

\begin{flushright}
BI-TP 2005/21\\
\end{flushright}

\vspace{\fill}

\begin{center}
{\Large\bf $J/\psi$ photo- and electroproduction, the saturation scale
and the gluon structure function}$^*$
\\[3.5ex]
{\bf Masaaki Kuroda}$^a$  and
{\bf Dieter Schildknecht$^b$ }\\[2.5mm]
$a$ Institute of Physics, Meiji Gakuin University \\[1.2mm]
Yokohama 244-8539, Japan \\[1.2mm]
$b$ Fakult\"{a}t f\"{u}r Physik, Universit\"{a}t Bielefeld \\[1.2mm] 
D-33501 Bielefeld, Germany \\[1.2mm]
and \\[1.2mm]
Max-Planck Institute f\"ur Physik (Werner-Heisenberg-Institut),\\[1.2mm]
F\"ohringer Ring 6, D-80805, M\"unchen, Germany
\end{center}

\vspace{\fill}

\begin{center}
{\bf Abstract}\\[2ex]
\begin{minipage}{\abstwidth}
Our approach to low-x deep inelastic scattering based on the $\vec l_\bot$
factorization of perturbative QCD (the color-dipole picture) yields 
parameter-free 
absolute predictions for $J/\psi$ production. The connection of $J/\psi$
production to the low-x saturation scale and to the gluon structure
function is clarified.
\end{minipage}
\end{center}

\vspace{\fill}
\noindent

\rule{60mm}{0.4mm}\vspace{0.1mm}

\noindent
${}^*$ Supported by Deutsche Forschungsgemeinschaft, contract number schi 
189/6-2 and the Ministry of Education and Science, Japan under the
Grant-in-Aid for basic research program B (no. 17340085)\\
\clearpage
\pagestyle{plain}
\setcounter{page}{1}

%%%%%\baselineskip 20pt

%\section{Introduction}

Vector meson photoproduction and electroproduction provide a significant
test of the theory of inclusive deep-inelastic scattering (DIS) at  
low values of the Bjorken scaling variable, $x \simeq Q^2/W^2 \ll 1$. In
the present paper, we confront our theoretical results \cite{Kuroda} 
on DIS and
vector-meson production with recently published experimental data on 
$J/\psi$ production \cite{Zeus}.

The low-x kinematics, within QCD, implies $\vec l_\perp$ 
factorization\footnote{Here $\vec l_\perp$ stands for the transverse
momentum of the gluon.} or, equivalently, the color-dipole picture 
\cite{Nikolaev}.
The  persistence  of the two-gluon-exchange structure\footnote{Note that
the structure of the amplitude, i.e. the expression for the color-dipole
cross section in transverse position space is dictated by the
gauge-invariant coupling of the two gluons to the $q \bar q$ pair.
This structure has to survive the transition to the ``soft'' domain
of $Q^2 \to 0$.} of the
$\gamma^*$-proton forward-scattering amplitude allows one to incorporate
low values of $Q^2$, including $Q^2 = 0$, into a unified description of 
the photoabsorption cross section at low x and all $Q^2$ \cite{Cvetic}. This
point  of view is supported by the empirical evidence for low-x
scaling \cite{DIFF2000,Cvetic}, $\sigma_{\gamma^*p} = \sigma_{\gamma^*p}
(\eta (W^2, Q^2))$,
that says that large and small values of $Q^2$ yield identical 
photoabsorption cross sections, once the corresponding energies are 
appropriately chosen to imply identical values of the scaling variable
$\eta (W^2, Q^2)$ that is explicitly defined below.

At low x, in the color-dipole picture from QCD, one may explicitly represent
the Compton-forward-scattering amplitude in terms of forward scattering of
$(q \bar q)^{J=1}$ (vector) states \cite{Ku-Schi}. 
In refs. \cite{Cvetic, Ku-Schi}, we made the
simplifying assumption that the forward-scattering amplitude is independent
of whether  the $(q \bar q)^{J=1}$ states have transverse or longitudinal
polarization. In the dual language of parton 
distributions,
in the kinematic domain where appropriate, this simplifying assumption
turned out to be equivalent to an underlying proportionality of the 
sea-quark and the gluon distribution \cite{Schi-Ku}, i.e.
\be
(q \bar q)_{\rm sea}~~ \sim~~ \alpha_s (Q^2) \cdot {\rm gluon~ distribution}
~~\sim~~ \Lambda^2_{sat} (W^2 = Q^2/x).
\label{1}
\ee
The proportionality (\ref{1}) together with the assumed power-like
increase of the ``saturation scale'', $\Lambda^2_{sat} (W^2)$, as a
function of the energy, $W$,
\be
\Lambda^2_{sat} (W^2) \sim (W^2)^{C_2},
\label{2}
\ee
upon requiring consistency with DGLAP evolution, led to the remarkable
conclusion that the value of $C_2$ must be fixed at \cite{Schi-Ku}
\be
C_2^{theory} = 0.276
\label{3}
\ee
in agreement with the previous fit \cite{Cvetic} to the experimental data,
\be
C_2^{exp} = 0.27 \pm 0.01.
\label{4}
\ee
The W-dependence of $\Lambda^2_{sat} (W^2)$ determines the approach to
saturation in the sense of
\be
\lim_{{W^2 \to \infty} \atop {Q^2 {\rm fixed}}} \frac{\sigma_{\gamma^*p}
\left( \eta (W^2, Q^2) \right)}{\sigma_{\gamma p} (W^2)} = 1,
\label{5}
\ee
or\footnote{Compare ref. \cite{HSQCD} for a plot of the experimental data
according to (\ref{5a}).}
\be
\lim_{{W^2 \to \infty} \atop {Q^2 {\rm fixed}}} 4 \pi^2 \alpha
\frac{F_2 (x, Q^2)}{\sigma_{\gamma p} (W^2)} = Q^2
\label{5a}
\ee
i.e. the approach to the ``soft'' energy dependence of the total
photoproduction cross section at any fixed value of $Q^2 > 0$. The
scaling variable in (\ref{5}) is given by
\be
\eta(W^2, Q^2) = \frac{Q^2 + m^2_0}{\Lambda^2_{sat} (W^2)}.
\label{6}
\ee
The mass $m^2_0 < m^2_\rho$ (where $m_\rho$ denotes the $\rho$-meson
mass) enters via quark-hadron duality \cite{Sakurai, Gorczyca}.

The saturation scale, $\Lambda^2_{sat} (W^2)$, specifies the effective
transverse momentum of the gluons coupled to the $q \bar q$ pair in the 
two-gluon exchange amplitude,
\footnote{Relation (\ref{7}) follows from
$\Lambda^2_{sat} (W^2) \equiv \frac{\pi}{\sigma^{(\infty)}} \int 
d \vec l_\perp^{~\prime 2} \vec l_\perp^{~\prime 2} 
\bar \sigma_{(q \bar q)^{J=1}_L} = \frac{6 \pi}{\sigma^{(\infty)}}
\int d \vec l_\perp^{~2} \vec l_\perp^{~2} \tilde \sigma 
(\vec l^2_\perp, W^2)$, where $\bar \sigma_{(q \bar q)^{J=1}_L}$
follows from $\tilde \sigma (\vec l_\bot^{~2}, W^2)$ by $J = 1$ projection,
and $\tilde \sigma (\vec l_\perp^{~2}, W^2)$
specifies what is
frequently called ``the unintegrated gluon distribution''. 
The above relation to be valid requires the normalization condition
$\frac{\sigma^{(\infty)}}{\pi} = \int d 
\vec l_\perp^{~2} \tilde \sigma (\vec l_\perp^{~2}, W^2) = \int d
\vec l_\perp^{~\prime 2} \bar \sigma_{(q \bar q)^{J=1}_L} 
(\vec l_\perp^{\prime 2}, W^2)$ to hold. The normalization 
condition must not necessarily hold, since $\tilde \sigma (\vec
l_\perp^{~2}, W^2)$ in general contains higher, $J>1$,
partial wave contributions.  
The above normalization condition is  true for  
$\tilde \sigma (\vec l_\perp^{~2}, W^2) = \frac{\sigma^{(\infty)}}{\pi}
\delta (\vec l_\perp^2 - \frac{1}{6} \Lambda^2_{sat} (W^2))$, that is
the ansatz for the dipole cross section
underlying our results for DIS and the present paper on $J/\Psi$
production.
Some (mild) deviation in the 
numerical factor of 1/6 in (\ref{7}) cannot be strictly excluded
purely on general grounds.}
\be
\langle \vec l^{~2}_\perp \rangle_{_{W^2}} = \frac{1}{6} \Lambda^2_{sat}
(W^2).
\label{7}
\ee
In addition to $\Lambda^2_{sat} (W^2)$, one needs the integral over the
transverse gluon distribution, $\sigma^{(\infty)}$,
that determines the normalization of the
total photoabsorption cross section. 
The specification of the two integrals over the unintegrated
gluon distribution, $\Lambda^2_{sat}(W^2)$ and $\sigma^{(\infty)}$,
is sufficient to determine the photoabsorption
cross section or the proton structure function at low $x$ for
$Q^2\ll \Lambda^2_{sat}(W^2)$ and $Q^2\gg \Lambda^2_{sat}(W^2)$,
respectively.  This is true under the above-mentioned 
assumption of the equality of transverse and longitudinal 
$(q\bar q)^{J=1}$ scattering or property (\ref{1}).
Compare also \cite{K-S}, where the equality is replaced
by a proportionality, and the connection with the longitudinal
and transverse parts of the proton structure function
is elaborated upon.  The complete dependence on $W$ and $Q^2$
for all $W$ and $Q^2$ at low $x$ not only requires
a knowledge of the integrated quantities but an ansatz for the
gluon-momentum dependence that specifies how 
$\Lambda^2_{sat}(W^2)$ and  $\sigma^{(\infty)}$ appear in the
photoabsorption cross section.

Since our representation of the virtual Compton-forward-scattering amplitude
explicitly contains the amplitudes for $(q \bar q)^{J=1}$ forward scattering,
the transition to vector-meson production is straight-forward indeed. 
Taking away the outgoing photon yields the production amplitude for a massive
$J=1$ quark-antiquark continuum. Integration of this continuum over an
appropriate mass interval, $\Delta M^2_V$, in the approximation of 
quark-hadron duality \cite{Sakurai}\footnote{Compare also ref. \cite{Martin}
for a recent application of quark-hadron duality.}, then determines the 
vector-meson-production cross
section. For $J/\psi$ production, in particular, we have \cite{Kuroda}
\bqa
& & \frac{d \sigma_{\gamma^*p \to J/\psi~ p}}{dt} (W^2, Q^2) \left|_{t=0} =
\right. \nonumber \\
& & \int_{\Delta M^2_{J/\psi}} dM^2 \int^{z_+}_{z_-} dz 
\frac{d \sigma_{\gamma^*p \to (c \bar c)^{J=1}p}}{dt~ dM^2~ dz} (W^2, Q^2, z, 
m^2_c, M^2),
\label{8}
\eqa
with
\be
z_\pm = \frac{1}{2} \pm \frac{1}{2} 
\sqrt{1-4 \frac{m^2_c}{M^2}},
\label{8a}
\ee
where $M^2 \equiv M^2_{c\bar c}$ denotes the mass of the produced $c \bar c$
pair. The mass of the charm quark, $m_c$, enters via the light-cone
wave function of the incoming virtual photon that describes the 
$\gamma^* \to c \bar c$ transition. We refer to ref. \cite{Kuroda} for
the explicit expression of the integrand in (\ref{8}).

The cross section on the right-hand side in (\ref{8}) depends on the values 
of $\sigma^{(\infty)}$ and on the parameters in $\Lambda^2_{sat} (W^2)$
as determined in our analysis of DIS. We have\footnote{Note 
that the relevant quantity in DIS is $R_{e^+e^-}
\sigma^{(\infty)}$.  The value of $\sigma^{(\infty)}$ in
(\ref{9}) corresponds to 4 quark flavors.}   \cite{Cvetic,Ku-Schi}
\be
\sigma^{(\infty)} = 48 GeV^{-2} = 18.7 mb,
\label{9}
\ee
and
\bqa
\Lambda^2_{sat} (W^2) & = & B \left( \frac{W^2}{W^2_0} + 1 \right)^{C_2}
\nonumber \\
& \cong & B^\prime \left( \frac{W^2}{1 GeV^2} \right)^{C_2},
\label{10}
\eqa
where $C_2$ is given by $C_2 = C_2^{theory}$ in (\ref{3}) and
\bqa
B & = & 2.24 \pm 0.43 GeV^2, \nonumber \\
W^2_0 & = & 1081 \pm 124 GeV^2,
\label{11}
\eqa
as well as
\be
B^\prime = 0.340 \pm 0.063 GeV^2.
\label{12}
\ee
The high-energy approximation in the second line of (\ref{10}) is even
satisfactory for low values of $W^2$. There are then essentially only two
adjusted quantities to describe DIS at low x, namely $\sigma^{(\infty)}$
and the normalization, $B^\prime$, of the saturation scale. At HERA energies,
we approximately have $2 GeV^2 \lsim \Lambda^2_{sat} (W^2) \lsim 7 GeV^2$.

The explicit expression for the cross section in (\ref{8}) in \cite{Kuroda}
ignores the finite (longitudinal) momentum transfer ($t_{min}\ne 0$)
occurring in the inelastic process of vector meson production
(in distinction from DIS), as well as the contribution
of  the real part of the amplitude to the $J/\Psi$-forward-production
cross section.  The finite momentum transfer in the
two-gluon-exchange approach implies different longitudinal
momenta ($x\ne x^\prime$) of the two gluons. These are
absent in the color-dipole or $\vec\ell_\perp$-factorization approach.
Their effect, called skewness, was analysed in terms of 
generalized gluon structure functions and can (approximately) be
incorporated by a multiplicative factor, i.e.
$\sigma^{(\infty)}$ in the expression for the $J/\Psi$-production in
(\ref{8}) is to be replaced by $\sigma^{(\infty)}R_g(C_2)$ 
\cite{Martin} with 
\bq
    R_g(C_2) = {{2^{2C_2+3}}\over{\sqrt\pi}}
        {{\Gamma(C_2+{5\over 2})}\over{\Gamma(C_2+4)}}
     \Bigg|_{C_2=0.276}   = 1.27,
\label{12a}
\eq
and $C_2$ being identical to the exponent of the $W$ dependence 
in (\ref{10}).
The correction for the real part of the production amplitude
for a power-law in energy $W$ approximately
amounts to a factor of $\sqrt{1+r^2}\cong 1.12$, yielding an increase
in the $J/\Psi$ production cross section by about 20\%
\cite{Martin}.  Altogether we thus have the substitution
\bq
    \sigma^{(\infty)} \to  \sigma^{\prime(\infty)}=
    \sigma^{(\infty)}\times 1.27\times 1.12 = 68.3 {\rm GeV}^{-2}
   \cong 26.6~ mb,
\label{12b}
\eq
to be applied to the explicit expression for the cross section
\cite{Kuroda} on the
right-hand side in (\ref{8}).

The $J/\psi$-production cross section (\ref{8}) in addition depends on
the charm-quark mass
\be
m_c = 1.5 GeV,
\label{13}
\ee
and on the integration interval,
\be
\Delta M^2_{J/\psi} = 3 GeV^2.
\label{14}
\ee
The integration over $dM^2$ in (\ref{8}) then runs over the mass
interval from $(2m_c)^2 = 3^2 GeV^2 = 9 GeV^2$ to 
$12 GeV^2$, where the upper integral boundary corresponds to
${1\over 2}(M(\Psi^\prime)^2+M(J/\Psi)^2)\cong 12 GeV^2$ with
$M(\Psi^\prime)=3.7 GeV$ and $M(J/\Psi)=3.1 GeV$.

The results of the experiments \cite{Zeus} were given in terms of the
$J/\psi$ production cross section, $\sigma_{\gamma^*p \to J/\psi~p} 
(W^2, Q^2)$,
and the $t$ distribution that was fitted by an exponential, $exp (-b 
\vert t \vert)$. Rather than attempting to extract the experimental forward
production cross section for $t \cong 0$ in (\ref{8}) from the experimental
data, we multiply the theoretical result (\ref{8}) 
including correction (\ref{12b}) by the inverse of the 
experimentally determined parameter $b$
\be
\sigma_{\gamma^*p \to J/\psi~ p} (W^2, Q^2) = \frac{1}{b}
\frac{d \sigma_{\gamma^*p \to J/\psi~ p}}{dt} (W^2, Q^2) \left|_{t=t_{{\rm min}}}
\right. \label{15}
\ee
and compare with the experimental results for $\sigma_{\gamma^*p \to J/\psi~p}
(W^2, Q^2)$. For $b$ we use a value of
\be
b = 4.5 GeV^{-2}
\label{16}
\ee
that is approximately equal to the experimental value for photoproduction
at $W = 90 GeV$. For $W \cong 30 GeV$ and $W \cong 300 GeV$, the parameter
$b$ decreases and increases, respectively, by approximately 8 \% and
17 \%. In electroproduction, $b$ decreases by about 10 \% to 20 \% at 
the largest available values of $Q^2$ \cite{Kiesling}

In fig. 1 and in fig. 2, we compare the theoretical predictions for the
$J/\psi$-production cross section according to (\ref{8}) 
(including corrections for skewing and the real part) and (\ref{15}) with
the experimental results from HERA. The parameters for the theoretical
predictions are specified in (\ref{9}) to (\ref{14}). There is 
agreement  with experiment for the $Q^2$ dependence at $W = 90 GeV$
shown in fig. 1, and for the $W$ dependence in photoproduction $(Q^2 = 0)$
in fig. 2.

The theoretical predictions in figs. 1 and 2 are based on the constant value
of $b = 4.5 GeV^{-2}$ from (\ref{16}). We have checked that the change 
of $b$ with $Q^2$ and with the energy, $W$, implies changes in the 
theoretical predictions that are within the error band of the experimental
data.

A comment on the reliability of our absolute predictions,
in particular on their normalization, may be appropriate.  
An increase of $\Delta M_{J/\Psi}^2= 3$ GeV$^2$ in (\ref{14}) to
$\Delta M_{J/\Psi}^2= 4$ GeV$^2$ or, alternatively,
a decrease of $m_c=1.5$ GeV to $m_c=1.4$ GeV in (\ref{13})
implies an increase of the $J/\Psi$ photoproduction cross section
by about 25 \%.  
%In our analysis, we have ignored the contribution
%of a real part in the amplitude to the 
%$J/\Psi$-production cross section.  For a power law, as in (\ref{2}), 
%in the imaginary part, the real part leads to an enhancement of the 
%cross section by about 20\%\cite{Martin}.
In connection with potential uncertainties of the absolute
normalization of the cross section in our approach it is worth
noting that other approaches \cite{Teubner}
require arbitrary fit factors ranging from 1.33 to 2.17
to achieve agreement with the data for $J/\Psi$ photoproduction.

With respect to an intuitive understanding of the above numerical results
from (\ref{8}), it will be rewarding to examine an approximate 
evaluation \cite{Kuroda}
of the cross section (\ref{8}). The approximation replaces the cross
section for the production of the $c \bar c$ open charm continuum 
on the right-hand
side in (\ref{8}) by its value at threshold,
\bqa
\frac{d \sigma_{\gamma^* p \to (c \bar c)^{J=1}p}}{dt~dM^2 dz}
\hspace*{-0.5cm} && (W^2, Q^2, z, m^2_c, M^2) \to \nonumber \\
&& \to \frac{d\sigma}{dt~dM^2~dz} \left(W^2, Q^2, z = \frac{1}{2},
M^2 = 4m^2_c = M^2_{J/\psi} \right).
\label{17}
\eqa
The integral in (\ref{8})
then reduces to
\be
\int^{4m^2_c+\Delta M^2_{J/\psi}}_{4 m^2_c} dM^2 \int^{z_+}_{z_-} dz =
\int^{4m^2_c+\Delta M^2_{J/\psi}}_{4 m^2_c} dM^2 \sqrt{1 - \frac{4m^2_c}{M^2}}
\equiv \Delta F^2(m^2_c, \Delta M^2_{J/\psi}).
\label{18}
\ee
For $M^2_{J/\psi}$ the
experimental value of $M^2_{J/\psi} = 3.1^2 GeV^2$ is substituted in 
(\ref{17}). It lies above threshold in the interval introduced via quark-hadron
duality,
\be
4 m^2_c < M^2_{J/\psi} < 4 m^2_c + \Delta M^2_{J/\psi}.
\label{19}
\ee
Quark confinement prevents the decay of the $J/\psi$ into free quarks. 

With
the approximation (\ref{17}), and introducing the notation from (\ref{18}),
the $J/\psi$-production cross section (\ref{8}) with correction 
(\ref{12b}) becomes
\bqa
&&\frac{d \sigma_{\gamma^*p \to J/\psi~p}}{dt} (W^2, Q^2) \left|_{t=t_{{\rm
        min}}\cong 0} 
\right. = 
\frac{3}{2} \frac{1}{16 \pi} \frac{\alpha~R^{(J/\psi)}}{3 \pi} 
(\sigma^{\prime(\infty)})^2 \cdot \nonumber \\
& \cdot & \frac{\Lambda^4_{sat}(W^2)}{(Q^2 + M^2_{J/\psi})^3}
\frac{1}{\left( 1 + \frac{\Lambda^2_{sat} (W^2)}{Q^2 + M^2_{J/\psi}}\right)^2}
\Delta F^2 (m^2_c, \Delta M^2_{J/\psi}),
\label{20}
\eqa
where $R^{(J/\psi)} = 4/3$. Expressing the electroproduction cross section
(\ref{20}) in terms of the
photoproduction cross section given by (\ref{20}) at $Q^2 = 0$,
\bqa
&&\frac{d \sigma_{\gamma p \to J/\psi~p}}{dt} (W^2, Q^2 = 0) \left|_{t=t_{{\rm min}}\cong 0}
\right. = \frac{3}{2} \frac{1}{16 \pi} \frac{\alpha R^{(J/\psi)}}{3 \pi}
(\sigma^{\prime(\infty)})^2 \cdot \nonumber \\
&& \cdot \frac{\Lambda^4_{sat} (W^2)}{\left( M^2_{J/\psi}\right)^3}
\frac{1}{\left( 1 + \frac{\Lambda^2_{sat} (W^2)}{M^2_{J/\psi}}\right)^2}
\Delta F^2 (m^2_c, \Delta M^2_{J/\psi}),
\label{21}
\eqa
we have
\bqa
&& \frac{d\sigma_{\gamma^*p \to J/\psi~ p}}{dt} (W^2, Q^2) \left|_{t=t_{
{\rm min}}\cong 0} 
\right. = \frac{d\sigma_{\gamma p \to J/\psi~ p}}{dt} (W^2, Q^2=0)
\left|_{t=t_{{\rm min}}\cong 0}
\right. \cdot \nonumber \\
&& \cdot \frac{M^2_{J/\psi}}{(Q^2 + M^2_{J/\psi})} \cdot
\frac{(M^2_{J/\psi} + \Lambda^2_{sat}(W^2))^2}{\left(Q^2 + M^2_{J/\psi} + 
\Lambda^2_{sat}(W^2)\right)^2}. 
\label{22}
\eqa
We stress that the strong increase (as $\Lambda^4_{sat} (W^2)$) of
$J/\psi$ photoproduction in (\ref{21}) is a unique consequence of the
threshold condition, $4m^2_c = 9 GeV^2 \simeq 3.1^2 GeV^2 = 9.6 GeV^2$. It
has nothing to do with the absolute value of $m^2_c$ and
$M^2_{J/\Psi}$ relative to $Q^2=0$.
At asymptotic energies, for $\Lambda^2_{sat} (W^2) \gg M^2_{J/\psi}$,
the $J/\psi$ photoproduction cross section in (\ref{21}) becomes
energy independent, or at most weakly dependent on energy, if we relax
the (approximate) constancy of $\sigma^{(\infty)}$ by allowing for a weak
dependence on $W$.

Since the $c \bar c$ mass in the cross section under the integral in
(\ref{8}) appears \cite{Kuroda}
in the combination of $Q^2 + M^2_{c \bar c}$, we
expect that the accuracy of the approximation of 
the $J/\psi$ forward-production cross section  given by (\ref{20}) will
improve with increasing $Q^2$. This is indeed seen in fig. 3. In fig. 3,
we also show the result obtained upon multiplication of (\ref{20}) by an
ad-hoc factor (of magnitude 2/3) that normalizes the cross section at
$Q^2 = 0$ to the empirical value of the photoproduction cross section. The
resulting $Q^2$ dependence, explicitly given in (\ref{22}), is consistent
with the experimental data.

The theoretical expression for the $Q^2$ dependence in (\ref{22}) that
follows from (\ref{8}) upon applying the approximation (\ref{17}) is of 
interest with respect to the fit of the $Q^2$ dependence by the ZEUS and
H1 collaborations. Their fit in terms of an ad hoc power-law ansatz,
\bqa
&& \sigma_{\gamma^*p \to J/\psi~ p} (W^2 = 90^2 GeV^2, Q^2) = \nonumber \\
&& = \sigma_{\gamma p \to J/\psi~ p} (W^2 = 90^2 GeV^2, Q^2 = 0)
\frac{\left( M^2_{J/\psi}\right)^n}{\left( Q^2 + M^2_{J/\psi}\right)^n}
\label{23}
\eqa
gave the result \cite{Kiesling}
\bqa
n & = & 2.486 \pm 0.080 \pm 0.068 \nonumber \\
& \cong &2.49 \pm 0.15.
\label{24}
\eqa
The success of this ad hoc fit is understood by comparison with our
theoretical result (\ref{22}). The additive contribution from
$\Lambda^2_{sat} (W^2)$ in the denominator of the theoretical expression
in (\ref{22}), in the fit with the ansatz (\ref{23}) is effectively 
simulated by the 
non-integral power of $n = 2.49$ given in (\ref{24}).

The H1 and ZEUS collaborations, by assuming s-channel 
helicity conservation, have extracted the ratio $R_{L/T}$ of 
longitudinal-to-transverse $J/\Psi$ production 
from their measurements of the $J/\Psi$ density-matrix elements.
The proportionality (\ref{1}) of the sea quark and the gluon
distribution or, equivalently,
the equality of forward production cross section for 
transverse and longitudinal polarization, with the 
approximation (\ref{17}), implies \cite{Kuroda}
\bq
     R_{L/T}= {{Q^2}\over{M_{J/\Psi}^2}}.
\label{24a}
\eq
A comparison with the experimental data\cite{Ciesielski, Kiesling}
for $R_{L/T}\cong r_{00}^{04}/(1-r_{00}^{04})$ shows approximate
agreement in the $Q^2$ dependence with a tendency for the absolute
normalization to  lie above the experimental result.

We turn to the interpretation of $J/\psi$ production in terms of the 
gluon-structure function. As a consequence from the duality of the color-dipole
picture, or $\vec l_\bot$ factorization, and $\gamma^*$-gluon scattering,
in the diffraction region of $x \ll 1$, and for $Q^2$ sufficiently large,
$Q^2 \gg \Lambda^2_{sat} (W^2)$, we have the identification 
\cite{Zakharov,Cvetic},
\be
\alpha_s (Q^2) x g(x, Q^2) = \frac{1}{8 \pi^2} \sigma^{(\infty)}
\Lambda^2_{sat} \left( W^2 = \frac{Q^2}{x} \right),
\label{25}
\ee
i.e. the function of $x$ and $Q^2$ on the lefthand side only depends on $W^2$,
once $x$ is replaced by $x = Q^2/W^2$. The identification (\ref{25}) holds
in the DGLAP region of $Q^2 \gg \Lambda^2_{sat} (W^2)$, 
where \cite{Cvetic, Schi-Ku}
\be
\sigma_{\gamma^*p} (\eta (W^2, Q^2)) \sim \frac{F_2 (x,Q^2)}{Q^2} \sim
\frac{\Lambda^2_{sat} (W^2)}{Q^2}.
\label{26}
\ee
We note that the factor $R_g(C_2)$ in (\ref{12a}) which is relevant for 
$J/\psi$ production is recovered 
by applying the substitution $x\to 0.41x$ \cite{ivanov} in
(\ref{25}).

The $W$-dependence of the saturation scale and its consequences with respect
to the gluon structure function are a unique result of our approach to DIS
at low x. According to (\ref{25}), a determination of $\Lambda^2_{sat} (W^2)$
by measuring the $W$-dependence of $J/\psi$ production according to 
(\ref{8}) or (\ref{20}) yields a unique x dependence of the gluon structure
function for any chosen fixed value of $Q^2 \gg \Lambda^2_{sat} (W^2)$. 
Since $\Lambda^2_{sat} (W^2)$ is independent of $Q^2$, it is irrelevant
at what value of $Q^2$ the energy dependence of $J/\psi$ production is
actually measured. A measurement of the energy dependence of e.g. $J/\psi$ 
photoproduction (at $Q^2 = 0$) yields the $x$-dependence of the gluon 
structure function for any 
fixed $Q^2 \gg \Lambda^2_{sat} (W^2)$ just as well as a measurement 
of the energy dependence of $J/\Psi$ production at
$Q^2 \gg \Lambda^2_{sat} (W^2)$.

Since the identification (\ref{25}) requires sufficiently large $Q^2$, a
representation of $J/\psi$ production as a function of $x$ and $Q^2$ in
terms of the gluon structure function can only exist for large values of
$Q^2$. Substituting (\ref{25}) into the large-$Q^2$ approximation of
(\ref{20}),
\bqa
\frac{d\sigma_{\gamma^* p \to J/\psi~p}}{dt} (W^2, Q^2) \left|_{t = t_{
{\rm min}}\cong 0}
\right. & = & \frac{3}{2} \frac{1}{16 \pi} \frac{\alpha R^{(J/\psi)}}{3 \pi}
\bigl(\sigma^{\prime(\infty)^2}\bigr) \cdot \nonumber \\
& \cdot & \frac{\Lambda^4_{sat} (W^2)}{\left( Q^2 + M^2_{J/\psi} \right)^3}
\Delta F^2 \left( m^2_c, \Delta M^2_{J/\psi}\right)
\label{27}
\eqa
we have
\bqa
&& \frac{d\sigma_{\gamma^* p \to J/\psi~p}}{dt} \left( W^2 = \frac{Q^2}{x},
Q^2\right) \biggl|_{t=t_{{\rm min}}\cong 0} \biggr. \nonumber \\
&& = \frac{3}{2} \frac{1}{16 \pi} \frac{\alpha R^{(J/\psi)}}{3 \pi}
\left( 8\pi^2 \right)^2 \cdot \frac{\left( \alpha_s(Q^2) x 
g(x,Q^2)\right)^2}{\left( Q^2 + M^2_{J/\psi} \right)^3} 
\cdot R_g^2(C_2)(1+r^2) \cdot \nonumber \\
&&~~~~\Delta F^2 \left( m^2_c, \Delta M^2_{J/\psi} \right),
\label{28}
\eqa
(where $Q^2 \gg \Lambda^2_{sat} (W^2)$)\footnote{The dependence on $(1/Q^6)
(\alpha_s(Q^2)xg(x,Q^2))^2$ in (\ref{28}) agrees with the one in ref.
\cite{Brodsky}}.

The notion of the gluon-structure function being used in the DGLAP fits of DIS
breaks down for small values of $Q^2$. For $Q^2 \to 0$ it becomes meaningless
to replace $\Lambda^2_{sat} (W^2)$ in the $Q^2 \rightarrow 0$ 
cross section in (\ref{21}) by the gluon structure function according
to (\ref{25}) with  
the aim of representing the $W$-dependence of
$J/\psi$ production in terms of the $x$-distribution of a
gluon structure function in the limit of
small $Q^2$, or $Q^2 \to 0$. As mentioned before, this does not prevent one
from measuring $\Lambda^2_{sat} (W^2)$ at $Q^2 = 0$ according to (\ref{21})
and from deducing the gluon structure function at $Q^2 \gg \Lambda^2_{sat}
(W^2)$ from such measurements.

An issue closely related to the above discussion concerns the prediction
of the energy dependence of $J/\psi$ photoproduction from the gluon-structure
function extracted from DIS measurements. According to the identification
(\ref{25}), the measured $x$-distribution of the gluon-structure function
at large $Q^2$ directly yields the W-dependence  of $J/\psi$ 
photoproduction by substitution of $\Lambda^2_{sat} (W^2)$ into (\ref{21}).

In fig. 4, in addition to the result from the numerical evaluation of 
quark-hadron duality from fig. 2, we show the result from the approximation
(\ref{21}) upon normalization by a factor of 3/2, as 
mentioned before in the discussion to fig. 3. The agreement 
with the more precise numerical evaluation of (\ref{8}) and with
the experimental results is very good
indeed. The fact that the $W$-dependence in the denominator in (\ref{21})
is relevant is seen by comparing with the result obtained upon ignoring
the denominator in (\ref{21}). The $W$-dependence becomes much too steep
and it even cannot be repaired by an ad hoc multiplication by a constant
factor. We note that ignoring the denominator in (\ref{21}) is equivalent
to incorrectly using the large-$Q^2$ approximation in (\ref{27}) and
(\ref{28}) at $Q^2 = 0$. This is of relevance with respect to the usual 
statement\footnote{Compare e.g. the talk by Teubner at DIS05
\cite{Teubner}} 
that the production of $J/\psi$ mesons even in photoproduction
is given by (\ref{28}) with the gluon structure function taken at an
appropriate scale. This conjecture is not supported by our analysis.
For $Q^2 \to 0$ a cross section of the form (\ref{21}) is relevant, where
$\Lambda^2_{sat}(W^2)$ may be identified with the gluon-structure function
at large $Q^2$ according to (\ref{25}).

Various fits of gluon structure functions from DIS have been 
used to predict \cite{Teubner}
$J/\psi$ photoproduction, some of them being successful after ad hoc
adjustments of the normalization by factors between 1.3 and 2.2. 
From our analysis, two conditions have
to be fulfilled for a successful representation of $J/\psi$ photoproduction:
\begin{itemize}
\item[i)] The gluon structure function at large $Q^2$ 
upon substitution of $x = Q^2/W^2$ has to 
fulfill (\ref{25}), at least
in good approximation. Otherwise, the right-hand side in (\ref{25}) will
depend on $W^2$ as well as $Q^2$, and a scale ambiguity will remain. 
In such a case no  unique conclusion on the
$W$ dependence of photoproduction can be obtained, 
since there is no preferred value
of $Q^2\gg \Lambda_{sat}(W^2)$ to be employed in the prediction 
for the energy dependence of $Q^2=0$ photoproduction.
\item[ii)] The $Q^2 \to 0$ cross section must be of the form (\ref{21}), 
where for $\bigl(\sigma^{(\infty)}\bigr)^2\Lambda^4_{sat} (W^2)$ 
in the numerator the large-$Q^2$ gluon-structure function
according to the proportionality (\ref{25}) is to be substituted. The form
of the cross section (\ref{21}) is a straight-forward consequence of
the underlying QCD structure, wherein two gluons of transverse momentum
$\vec l_\perp$  couple to the $c \bar c$ pair,
combined with the massive-quark threshold relation, $M^2_{J/\psi} \cong
4 m^2_c$. Ignoring the $W^2$-dependent denominator in (\ref{21}) corresponds
to incorrectly applying the large-$Q^2$ form (\ref{28}) at $Q^2 = 0$
by putting $Q^2 = 0$ in the denominator.
\end{itemize}
In summary, based on the coupling of the $c \bar c$ pair to two gluons
according to perturbative QCD,
with $\sigma^{(\infty)}$ and $\Lambda^2_{sat} (W^2)$ taken from the
analysis of the total photoabsorption cross section, we have obtained
an absolute prediction of forward $J/\psi$ photo- and electroproduction.
The successful application of quark-hadron duality implies that the 
dependence of the cross section on the wave function of the outgoing
$J/\psi$ meson can be neglected. The final results can be put into a very
simple form that allows for a transparent understanding of the underlying
theoretical ansatz.

\newpage

\begin{figure}[htbp]\centering
\epsfysize=14cm
\centering{\epsffile{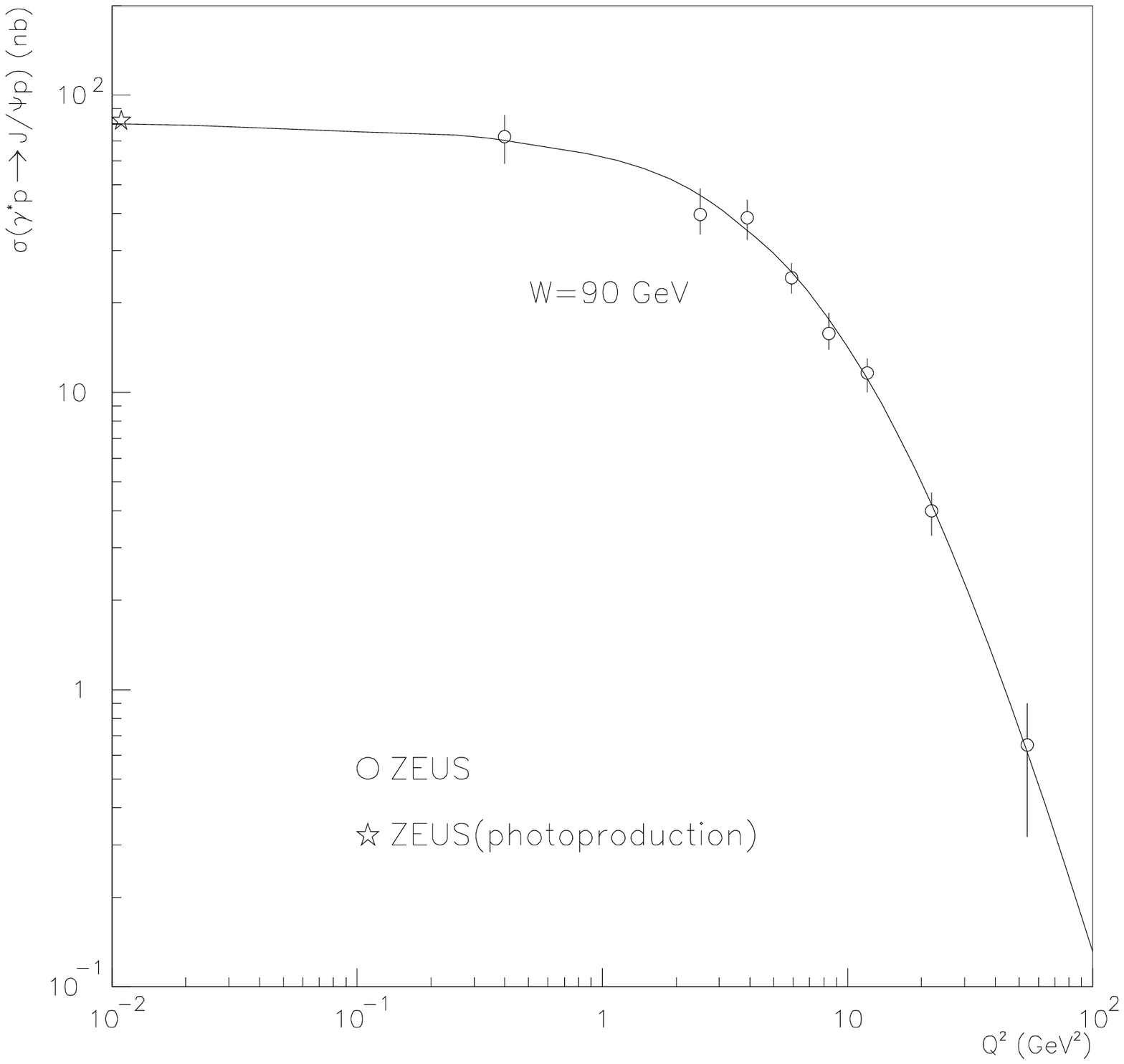}}
\caption{The $Q^2$-dependence of the cross section for $J/\psi$ 
production, $\sigma_{\gamma^*p \to J/\psi~ p}(W^2, Q^2)$, at the energy $W$ of 
$W = 90 GeV$. The theoretical curve is obtained by applying 
charm-quark hadron duality to $\gamma^*p \to c \bar c~p$ forward production. 
The experimental data are from the ZEUS collaboration \cite{Zeus}.}
\label{fig1}
\end{figure}

\begin{figure}[htbp]\centering
\epsfysize=14cm
\centering{\epsffile{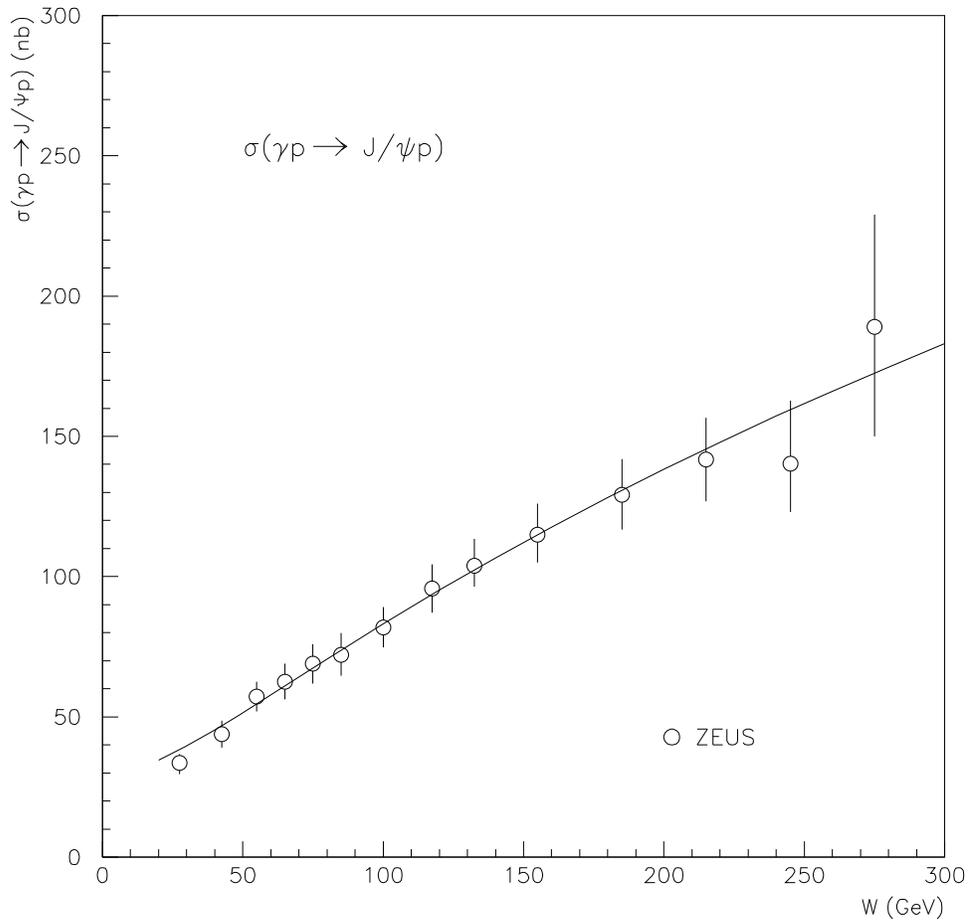}}
\caption{Same as fig. 1, but for the $W$-dependence of $J/\psi$
photoproduction $(Q^2 = 0)$.}
\label{fig2}
\end{figure}

\begin{figure}[htbp]\centering
\epsfysize=14cm
\centering{\epsffile{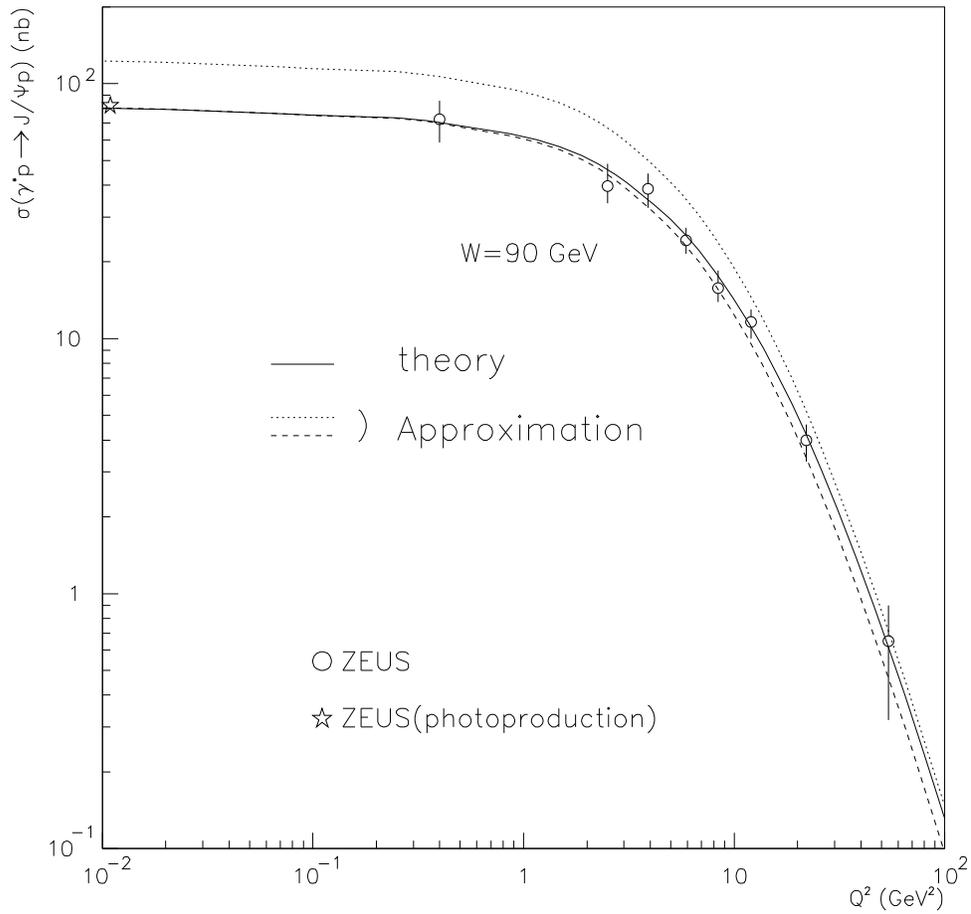}}
\caption{In addition to the results in fig. 1, we show the 
$Q^2$-dependence according to (\ref{20}), obtained by the approximate 
evaluation of charm-quark hadron duality (dotted curve). The lower 
(dashed) curve is obtained
by normalizing at $Q^2 = 0$ to photoproduction (compare also (\ref{22}))
by multiplication of the
result in (\ref{20}) and (\ref{21}) with an appropriate factor.}
\label{fig3}
\end{figure}

\begin{figure}[htbp]\centering
\epsfysize=14cm
\centering{\epsffile{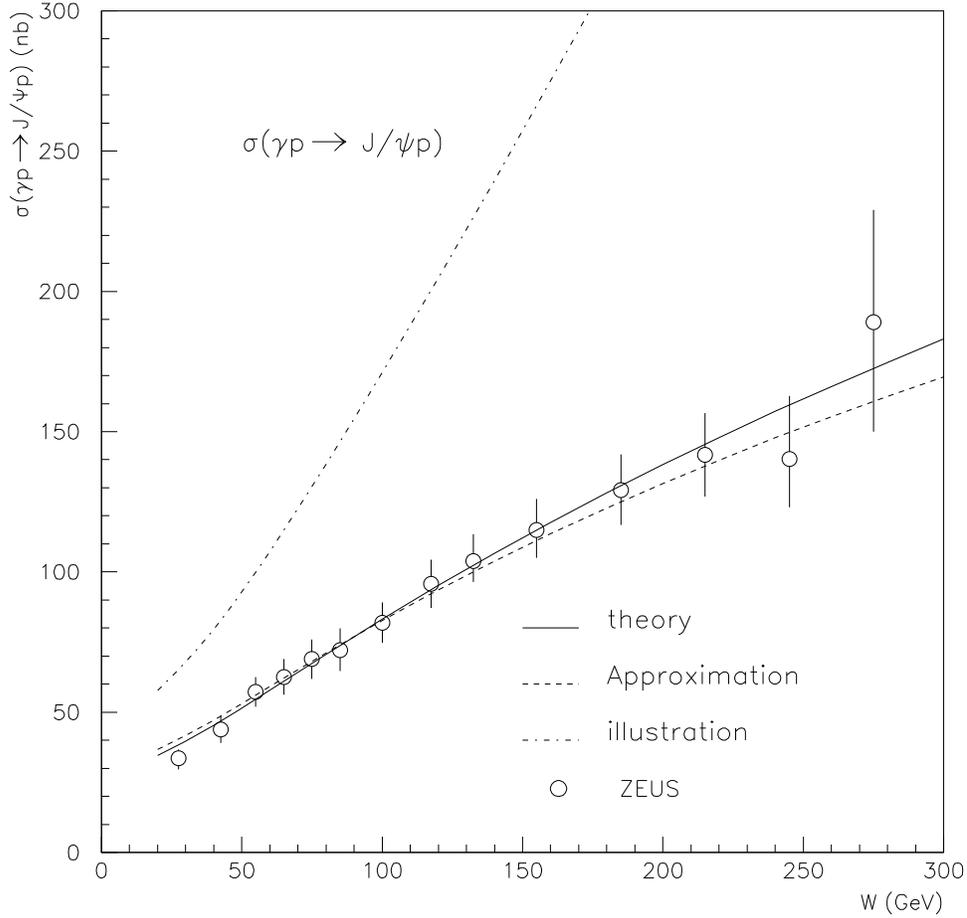}}
\caption{In addition to the results in fig. 2, we show the 
$W$-dependence from the approximation (\ref{21}) normalized to the 
experimental result at $W = 90 GeV$ as in fig. 3 (dashed curve). 
The dash-dotted 
curve 
illustrates what happens, if the large-$Q^2$ approximation in (\ref{27})
and (\ref{28}) is -- incorrectly -- applied by putting $Q^2 = 0$ in the
denominator i.e. by ignoring the $W$-dependent factor in the denominator of
(\ref{21}).}
\label{fig4}
\end{figure}
\end{document}